\documentclass[12pt]{elsart}
\usepackage{graphicx}
\usepackage{amssymb}
\usepackage{times}
\begin{document}

\newcommand{\re}{\mathop{\mathrm{Re}}}
\newcommand{\im}{\mathop{\mathrm{Im}}}
\newcommand{\D}{\mathop{\mathrm{d}}}
\newcommand{\I}{\mathop{\mathrm{i}}}

\def\lambar{\lambda \hspace*{-5pt}{\rule [5pt]{4pt}{0.3pt}} \hspace*{1pt}}

\noindent {\Large DESY 11-040}

\noindent {\Large March 2011}

\bigskip

\begin{frontmatter}

\journal{Phys. Rev. ST-AB}

\date{}

\title{Generation of attosecond soft x-ray pulses in a longitudinal space charge amplifier}

\author{M.~Dohlus},
\author{E.A.~Schneidmiller}
\author{and M.V.~Yurkov}

\address{Deutsches Elektronen-Synchrotron (DESY),
Notkestrasse 85, D-22607 Hamburg, Germany}

\begin{abstract}
A longitudinal space charge amplifier (LSCA), operating in soft x-ray regime, was recently proposed.
Such an amplifier consists of a few amplification cascades (focusing channel and chicane) and a short radiator
undulator in the end. Broadband nature of LSCA supports generation of few-cycle pulses as well as wavelength
compression. In this paper we consider an application of these properties of LSCA for generation of attosecond
x-ray pulses. It is shown that a compact and cheap addition to the soft x-ray
free electron laser facility FLASH would allow to generate 60 attosecond (FWHM) long x-ray pulses with
the peak power at 100 MW level and a contrast above 98\%.
\end{abstract}

\end{frontmatter}

\bigskip

\baselineskip 20pt

\clearpage

\section{Introduction}

Longitudinal space charge (LSC) driven microbunching instability \cite{fel-bc,lsc-mb} in electron linacs with
bunch compressors (used as drivers of short wavelength FELs) was a subject of intense theoretical and experimental
studies during the past years \cite{huang-2004,venturini,huang-chao,loos,schmidt,lumpkin,clemens,heater-oper}.
Such
instability develops in infrared and visible wavelength ranges and can hamper electron beam diagnostics and
free electron laser (FEL) operation.

It was proposed in \cite{lsca-prst} to use this effect for generation of vacuum ultraviolet (VUV) and
x-ray radiation.
A concept of a longitudinal space charge amplifier (LSCA) was introduced,
scaling relations for an optimized LSCA were obtained, and its possible applications were analyzed.
It was pointed out, in particular, that a broadband nature of LSCA supports generation of attosecond pulses.

There are many different schemes for generation of attosecond X-ray pulses from free electron lasers (FELs)
\cite{attofel-oc,oc-2004-2,atto-b,atto-e,atto-f,prstab-2006-2,huang-24as}. Most of the proposed schemes make
use of a short intense laser pulse to manipulate electron energy in a small slice of an electron bunch and to
select only this short slice with the help of FEL mechanism.

In Ref.~\cite{lsca-prst} a scheme was suggested that uses
the longitudinal space charge amplifier in combination with laser manipulation of the electron
beam to produce attosecond soft X-ray pulses.
In this paper we describe the operation of the LSCA-based attosecond scheme in details,
exemplify it with the parameters of the soft X-ray
free electron laser facility FLASH \cite{flash-nat-phot,njp}, perform thorough numerical simulations,
and discuss a possible implementation at FLASH making use of the existing infrastructure.

\section{Operation of the longitudinal space charge amplifier}

\begin{figure*}[tb]

\includegraphics[width=1.0\textwidth]{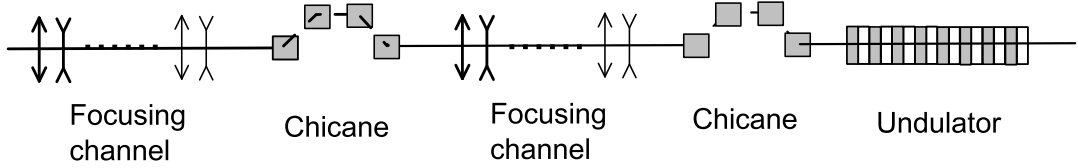}

\caption{\small Conceptual scheme of an LSC amplifier.}

\label{scheme}
\end{figure*}

The scheme (see Fig.~\ref{scheme}) of the longitudinal space charge amplifier (LSCA) \cite{lsca-prst}
is simple both conceptually and technically.
An amplification cascade consists of a focusing
channel and a dispersive element (usually a chicane) with an optimized momentum compaction $R_{56}$.
In a channel the energy modulations are accumulated, that are proportional to density modulations and
space charge impedance of the drift space. In the chicane these energy modulations are converted into
induced density modulations that are much larger than initial ones \cite{fel-bc}, the ratio defines a gain per
cascade. In this paper we will consider the case when the amplification starts up from the shot noise in
the electron beam. A number of cascades is defined by the condition that the total gain, given by
the product of partial gains in each cascade, is sufficient
for saturation (density modulation on the order of unity) after the start up from shot noise.
The amplified density modulation has
a large relative bandwidth, typically in the range 50-100 \%.
Behind the last cascade a radiator undulator is installed, which
produces a powerful radiation with a relatively narrow line
(inverse number of periods) within the central cone. This radiation is transversely coherent,
and the longitudinal
coherence length is given by the product of the number of undulator periods by the radiation wavelength.
When LSCA saturates in the last cascade, a typical enhancement of the radiation power
over that of spontaneous emission is given by a number of electrons per wavelength.

\section{Description of the LSCA-based attosecond scheme}

First few cascades of LSCA (as in Fig.~1) operate as described above.
A broadband density modulation is amplified around the
optimal wavelength

\begin{equation}
\lambda_{0} \simeq 2\pi \sigma_{\bot}/\gamma ,
\label{opt-wl}
\end{equation}

\noindent where $\sigma_{\bot}$ is a transverse size of the beam,
and $\gamma$ is relativistic factor. Such a choice of the wavelength is optimal because the LSC impedance
reaches its maximum, and, at the same time, transverse correlations of the LSC field
are on the order of the beam size \cite{venturini},
what guarantees a good transverse coherence of the radiation in the end.
Length of a drift space\footnote{By a "drift space" or a "drift" we
mean in this paper a focusing channel, in which LSC effect accumulates. It is a drift from a point of view of
longitudinal dynamics, but not a transverse one.}, beta-function, momentum compaction
$R_{56}$ of a chicane are optimized for the chosen wavelength.

\begin{figure*}[tb]

\includegraphics[width=1.\textwidth]{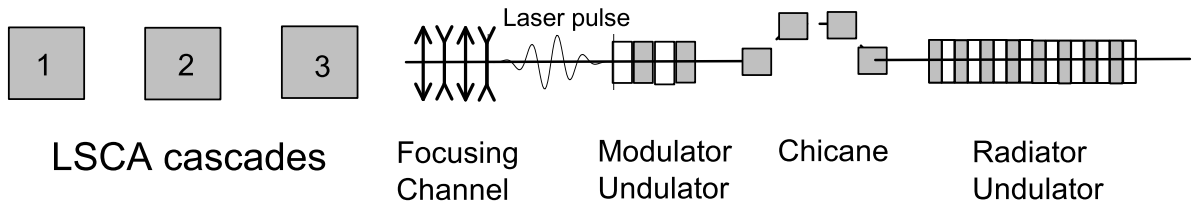}

\caption{\small LSCA-based attosecond scheme. Numbered boxes in the left part of the figure denote standard
cascades of LSCA (focusing channel plus chicane).}

\label{atto-scheme}
\end{figure*}

\begin{figure*}[tb]

\includegraphics[width=0.49\textwidth]{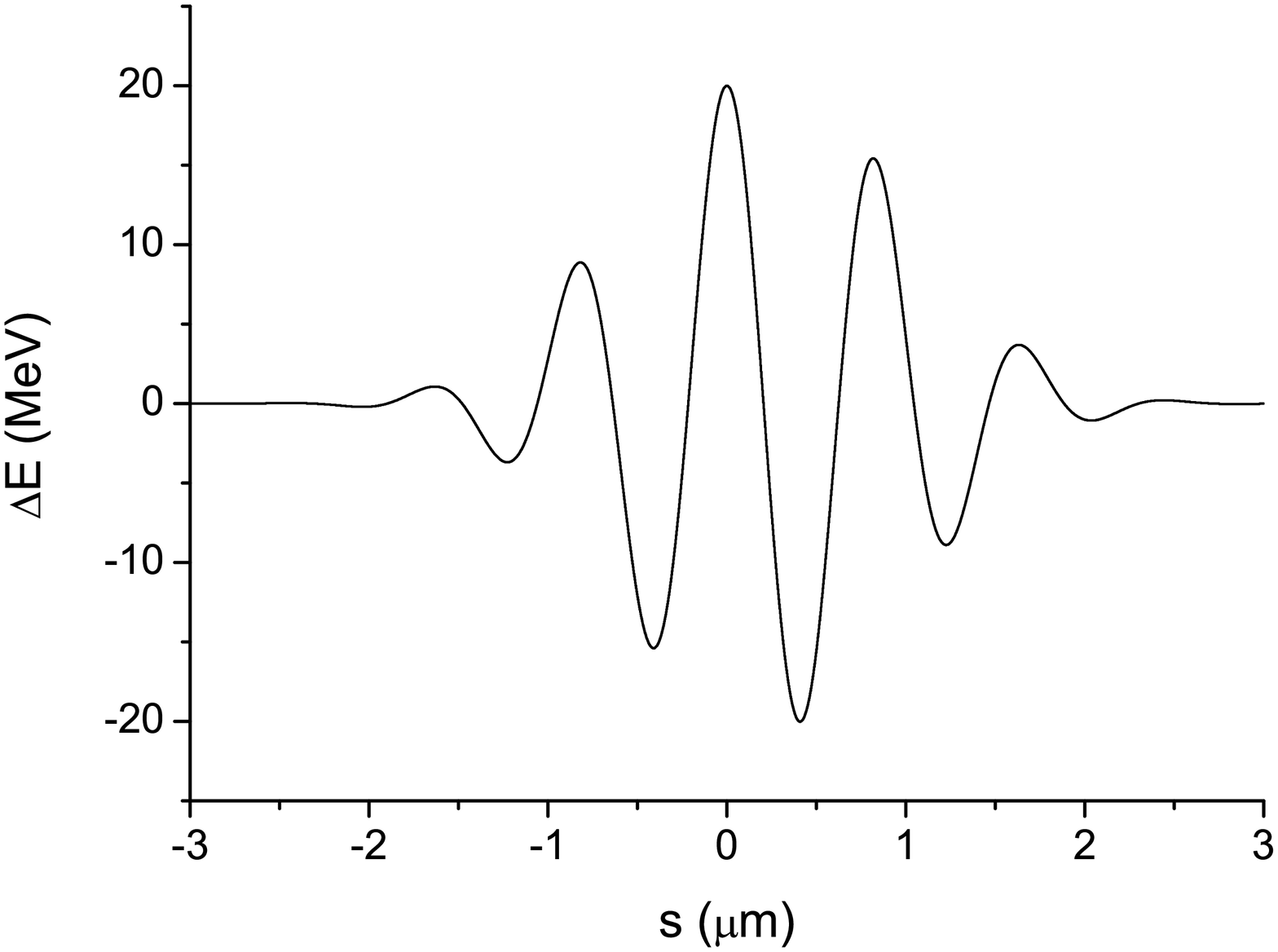}

\caption{\small Energy modulation, induced on the beam by a short laser pulse in a two-period undulator.
Energy deviation is plotted as a function of the longitudinal position in an electron bunch.}

\label{atto-mod}
\end{figure*}

The last cascade is modified as shown in Fig.~2. A short two-period modulator undulator is installed in
front of the last chicane. In this undulator the electron beam is modulated in energy (modulation wavelength is
much longer than $\lambda_{0}$) by a few-cycle powerful
laser pulse\footnote{Note that the laser beam can be in-coupled either through the previous chicane
where the electron beam has a transverse offset, or through some bending system upstream of the first
cascade of LSCA.}
in the same way as it is suggested for FEL-based attosecond
schemes \cite{oc-2004-2,atto-b,atto-f,prstab-2006-2,huang-24as}, see Fig.~\ref{atto-mod} as an example.
A short slice in the electron bunch
(between 0 and 0.3 $\mu$m in Fig.~\ref{atto-mod}) gets the
strongest energy chirp, and is being strongly compressed (by a factor $C \gg 1$)
in the following chicane. The $R_{56}$ of this chicane
and the chirp are adjusted such that a required wavelength compression within that slice is achieved, and, at the
same time the amplification of the microbunching within the slice
through the last chicane is optimal (parameters of the whole
amplification chain are adjusted such that saturation is reached in the last chicane). Other parts of the
electron bunch are either uncompressed or have much weaker compression than a slice with the strongest
energy chirp.

The chicane is followed by a short radiator undulator of which resonance wavelength,
$\lambda_{0}/C$, corresponds to the
(broadband) wavelength spectrum within the strongly compressed slice. Then only this short slice of the electron bunch
produces undulator radiation within the central cone (which is selected by a pinhole downstream),
while modulation wavelengths in the rest of the electron bunch are much longer than the resonance wavelength of
the undulator. In other words, the rest of the electron bunch produces only spontaneous emission within the
central cone, and this emission would define the contrast of generated attosecond pulses.

Note that the last
part of the scheme (modulator undulator, chicane, and radiator undulator) operates in a similar way as the last
part of the attosecond scheme proposed in Ref.~\cite{huang-24as}.
The two schemes differ essentially by mechanisms of
the generation of
microbunching before wavelength compression. The authors of \cite{huang-24as} considered echo-enabled harmonic
generation \cite{eehg}, while our scheme is based on the longitudinal space charge amplifier, described above.

Finally, let us note that broadband nature of LSCA makes wavelength compression especially
attractive \cite{lsca-prst}. Indeed, the compression factor is given by the formula:

\begin{equation}
C = (1-hR_{56})^{-1} \ ,
\label{comp-fac}
\end{equation}

\noindent where $h$ is the linear energy chirp (the derivative of relative energy deviation). For a large $C$ a
variation of the compression factor reads:

\begin{equation}
\frac{\Delta C}{C} \simeq C \ \frac{\Delta h}{h} \ .
\label{comp-del}
\end{equation}

\noindent After the compression the bands of density modulations and of the radiator must overlap. This leads to
the following requirement on the compression stability:

\begin{equation}
\frac{\Delta C}{C} < \frac{\Delta k_{max}}{k} \ ,
\label{comp-del-1}
\end{equation}

\noindent where $\Delta k_{max}= max(\Delta k_{den},\Delta k_{rad})$, and $\Delta k_{den}$ and $\Delta k_{rad}$
are bandwidths of the density modulation and of the radiator, respectively. Thus, the stability of the chirp
must satisfy the requirement:

\begin{equation}
\frac{\Delta h}{h} < \frac{1}{C} \ \frac{\Delta k_{max}}{k} \ .
\label{h-del}
\end{equation}

For coherent FEL-type modulations and an undulator as a radiator $\Delta k_{max}/k \ll 1$ what might set very tight
tolerance for the chirp stability and limit practically achievable compression factors. For an LSCA, however,
$\Delta k_{max}/k = \Delta k_{den}/k \simeq 1$, so that for a given chirp stability one can go for much
stronger compression.
Alternatively, for a given compression factor one can significantly loosen the tolerances. Note also that
nonlinearities of the longitudinal phase space do not play a significant role in the case of LSCA.

\section{Parameter set for an attosecond LSCA}

We exemplify an operation of the attosecond scheme with the parameters of the soft x-ray FEL facility
FLASH \cite{flash-nat-phot,njp}. In the Section 6 we will discuss a possible technical implementation of this
scheme making use of the existing infrastructure at FLASH.

Concerning the choice of parameters of the electron beam, we rely on the results of beam dynamics simulations
\cite{dohlus-zagor,zagor}.
For the case of the bunch charge 100 pC we reduce compression by a factor of two (with respect to the
scenario considered in \cite{zagor}), thus ending up with the slice parameters presented in Table~1.

\begin{table}[tb]
\caption{Electron beam parameters}

\begin{tabular}{l l}
\hline
Energy  &  1.2 GeV \\
Charge  &  100 pC  \\
Peak current  &  1 kA \\
Slice energy spread  &  150 keV \\
Slice emittance & 0.4 mm mrad \\
\hline
\end{tabular}
\end{table}

Number of LSCA cascades, their parameters, and the operating wavelength range are chosen with the help of the
guidelines of Ref.~\cite{lsca-prst}. As a focusing structure we choose FODO-lattice with the period
1.4 m and the average beta-function also 1.4 m (even smaller beta-function would be preferable,
but the technical feasibility and operational issues
have to be taken into account). For electron beam parameters from Table 1, we calculate from
(\ref{opt-wl}) that the optimal wavelength for amplification in LSCA is around 40 nm.
The length of a drift in each cascade is chosen
to be 2.8 m, or two FODO periods. This length should provide a sufficient gain per cascade, it is smaller
than a reduced wavelength of plasma oscillations, and the longitudinal velocity spread due to emittance does not
play any role \cite{lsca-prst}. The optimal $R_{56}$ of the chicane for a given energy spread and
wavelength range is about 50 $ \mu$m.
The chicane in each cascade should fit in a space between two quadrupoles.
Thus, the total length of a cascade is about 3.5 m (or 2.5 FODO periods).

According to our estimates, we need three regular cascades with the parameters, described above, and
a last special cascade (see Fig.~\ref{atto-scheme}), so that a total length of the system is about 14 m.
In the last cascade we have a drift of the same length (2.8 m),
followed by a two-period undulator and a chicane with reduced $R_{56}$.
The undulator has a period length of 10 cm and a peak field of 1.4 T,
so that a resonance of 1.2 GeV beam with the laser beam (wavelength is 800 nm) is provided.
In the last cascade a final amplification and, at the same time, a wavelength compression (down to 4-5 nm)
in a short slice
take place. To estimate the required energy chirp and the $R_{56}$ of the last chicane, we have to take
the following considerations into account.
The amplitude of density modulation at the
compressed wavelength should not be strongly suppressed by the uncorrelated energy spread
during compression. For a Gaussian
energy distribution an amplitude gain per cascade is given by \cite{fel-bc}:

\begin{equation}
G = Ck_0|R_{56}|
\frac{I}{\gamma I_{\mathrm{A}}}
\frac{4\pi|Z(k_0)| L_d}{Z_0}
\exp \left(-\frac{1}{2}C^2k_0^2R_{56}^2\frac{\sigma_{\gamma}^2}{\gamma^2}
\right) \ .
\label{gain-wake}
\end{equation}

\noindent Here $k_0= 2\pi/\lambda_0$ is the modulation wavenumber before compression,
$Z$ is the impedance of a drift space (per unit length),
$Z_0$ is the free-space impedance, $L_d$ is the length of the drift space,
$I$ is the beam current,
$I_{\mathrm{A}}$ is the Alfven current, $\gamma$ is
relativistic factor, and $\sigma_{\gamma}$ is rms uncorrelated energy
spread (in units of rest energy). The beam current and energy spread are taken before compression.
The optimal $R_{56}$ from (\ref{gain-wake}) is given by
$R_{56} \simeq \gamma (\sigma_{\gamma} k_f)^{-1}$, where $k_f = C k_0$ is the modulation wavenumber
after compression. Then the required energy chirp can be defined from (\ref{comp-fac}):

\begin{equation}
h = \frac{C-1}{CR_{56}} \simeq \frac{C-1}{C} \ \frac{\sigma_{\gamma}}{\gamma} \ k_f \ .
\label{chirp-req}
\end{equation}

\noindent If the electron beam is modulated by a laser with the wavelength $\lambda_L = 2 \pi /k_L$, and the
modulation amplitude is $\delta \gamma$, then the chirp at zero crossing is
$h \simeq k_L \delta \gamma/ \gamma$. Thus, with the help of (\ref{chirp-req}) we get an estimate of a required
amplitude of energy modulation:

\begin{equation}
\delta \gamma \simeq \frac{C-1}{C} \ \frac{\lambda_L}{\lambda_f} \ \sigma_{\gamma} \ .
\label{en-mod}
\end{equation}

\noindent If the compression factor is about 8-10, $\lambda_L = 800$ nm, $\lambda_f \simeq 5$ nm, then
the required energy modulation amplitude is about 20 MeV, and the corresponding $R_{56}$ is 6-7 $\mu$m.
We should note that formula (\ref{gain-wake}) is valid for a linear regime of amplification, while we
deal with a
saturation in the last cascade. Moreover, energy modulations due to a short laser pulse are not sinusoidal.
Therefore, the above obtained estimate can only be used as a first guess.
For our simulations we choose the modulation amplitude of 20 MeV, and the
$R_{56}$ was optimized in the simulations at 7.1 $\mu$m.
In order to get the desired energy modulations, we use a laser pulse with the
duration of 5 fs (FWHM) and pulse energy 3 mJ. The laser beam is focused into the modulator undulator,
the spot size in the waist is $w_0 = 300 \ \mu$m.

Finally, we should choose parameters of the radiator undulator that should be tuned to the desired wavelength
of 5 nm for the given beam energy of 1.2 GeV. We take a planar undulator with the
following parameters: undulator period is 2.5 cm, peak field is 0.67 T, number of periods is 5.
We assume that the undulator has a
tunable gap which would allow us to change wavelength by
changing the gap and/or the beam energy.

\section{Numerical simulations of the attosecond LSCA}

\begin{figure*}[tb]

\includegraphics[width=0.49\textwidth]{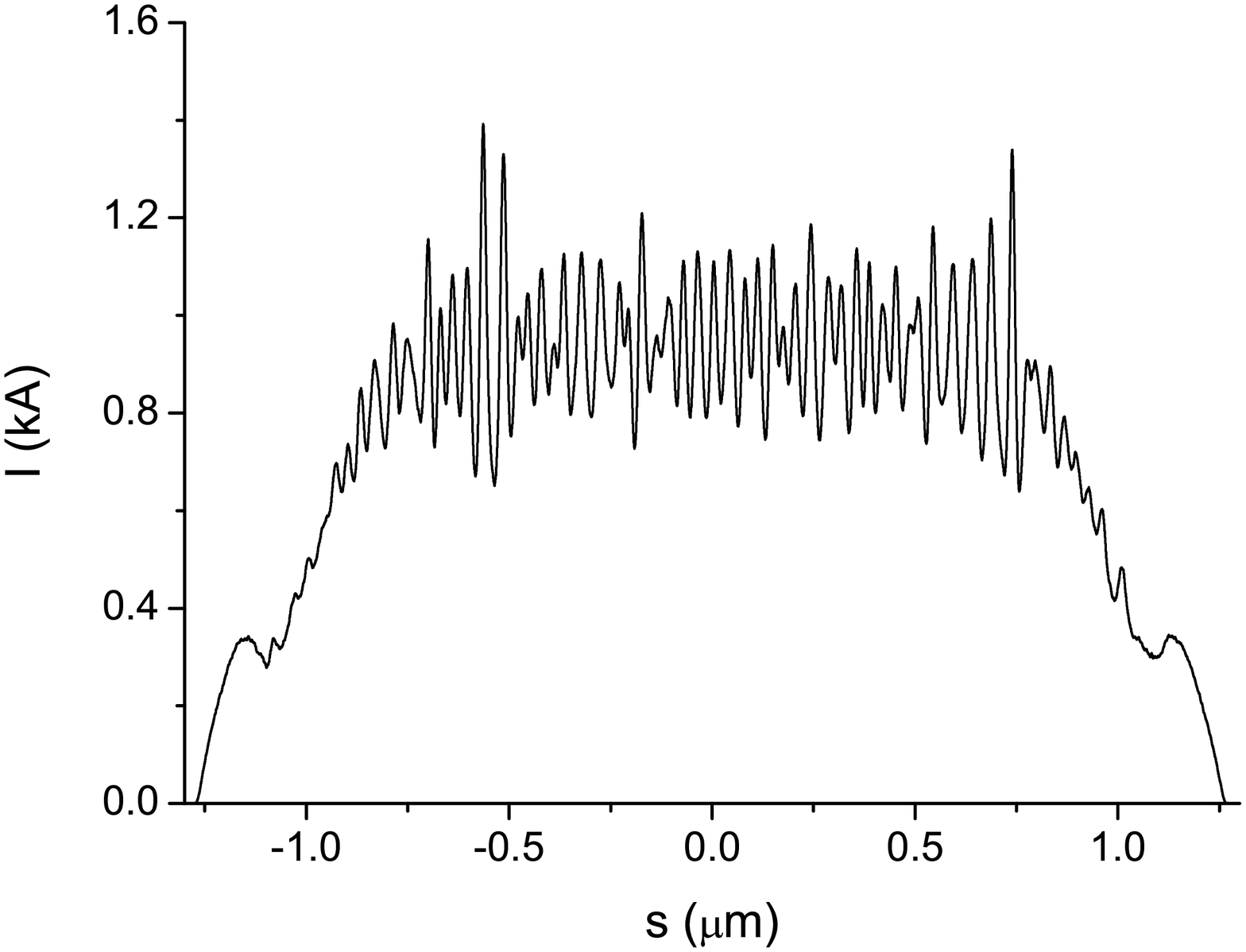}
\includegraphics[width=0.49\textwidth]{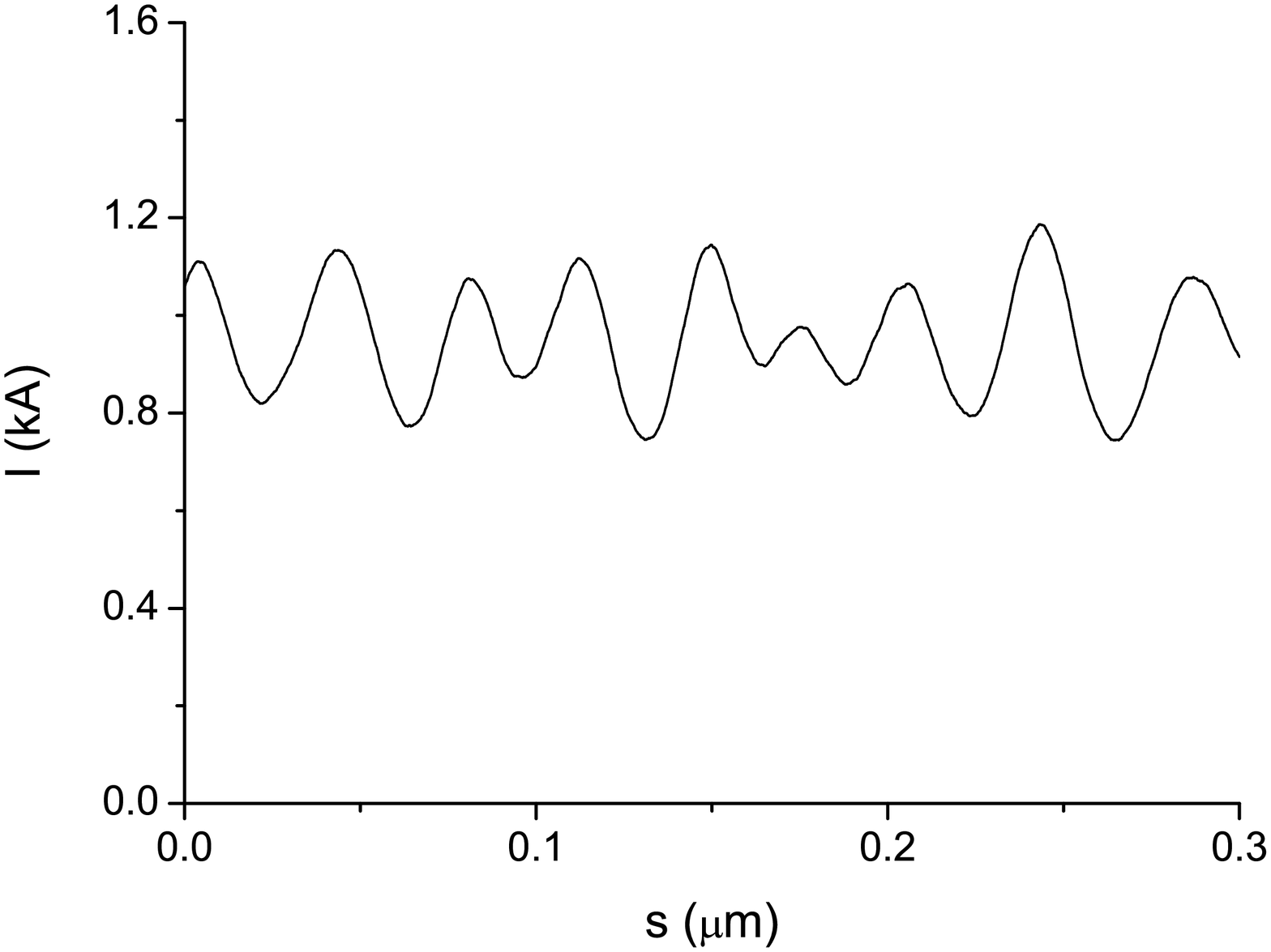}

\caption{\small  Modulated electron bunch (left) and its enlarged fraction (right) after the 3rd amplification
cascade. The calculations were performed with Astra.}

\label{fig:astra-time}
\end{figure*}

\begin{figure*}[tb]

\includegraphics[width= .49 \textwidth]{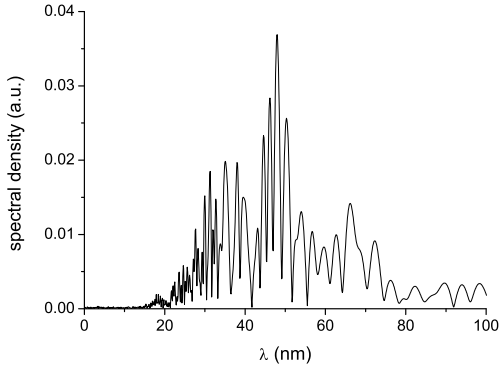}

\caption{\small Spectrum of the modulated electron current (modulus of the bunch form factor)
after the 3rd amplification cascade. The calculations were performed with Astra.}

\label{fig:astra-spectr}
\end{figure*}

The numerical simulations of the LSCA operation were done as follows. Self-interaction of electrons
in the beam by space charge fields in the focussing channels was simulated with 3-D version of the space charge
tracking code Astra \cite{astra,astra-3d}.
All important aspects of the problem like betatron motion of particles in the FODO channel,
three-dimensional calculation of
the space charge field, start up from shot noise were included in the simulations. To properly
treat the start up from noise, we took only a short part of the bunch with the length about 2 $\mu$m,
and used a real number of particles, $4\times 10^7$, distributing them randomly in 6-D phase space.
After the energy change of each particle in a drift was calculated, we simply applied $R_{56}$ to add
the effect of a chicane (later we checked that CSR effects do not play any significant role, see below).
In this way the evolution of the particle distribution through a single cascade was
simulated. Then the procedure was repeated for the next cascade, etc.
The results of simulations of the first three cascades of LSCA are
shown in Figs.~\ref{fig:astra-time} and \ref{fig:astra-spectr}. In Fig.~\ref{fig:astra-time}
we show the amplified noisy modulations of beam current, and in Fig.~\ref{fig:astra-spectr}
the modulus of the bunch form factor (Fourier transform of the trace in Fig.~\ref{fig:astra-time}, left)
is presented. Note that
a linear gain (in amplitude) per cascade is about 5 at 40-50 nm wavelength. In the third cascade the
amplification becomes slightly nonlinear, so that the gain is reduced by about 10\%. Typical feature size
in spectrum (interval of spectral coherence) can be estimated as $\Delta \omega \simeq 2\pi c / l_b$ \cite{book},
where $l_b \simeq 2 \ \mu$m is the
length of a part of the bunch in our simulations. Converting this dependence to a wavelength interval, we
get $\Delta \lambda \simeq \lambda^2 / l_b$, i.e. one should
expect a quadratic dependence of the coherence interval on the wavelength. Also note that in case of
simulations with Astra we
used only a short part of a 100 pC bunch, so that the fine structure in spectrum does not correspond to
a real situation with much longer bunch.

\begin{figure*}[tb]

\includegraphics[width=0.49\textwidth]{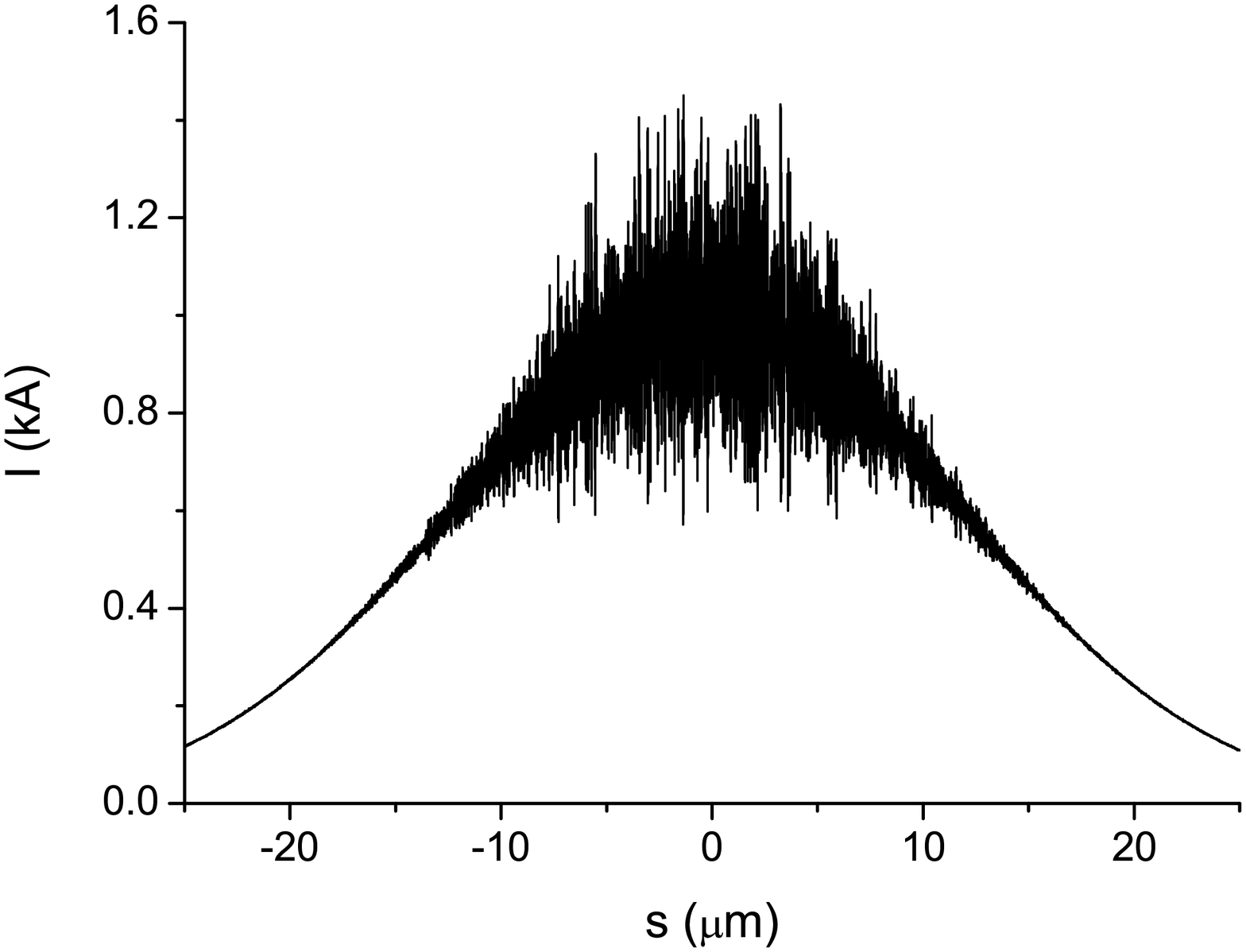}
\includegraphics[width=0.49\textwidth]{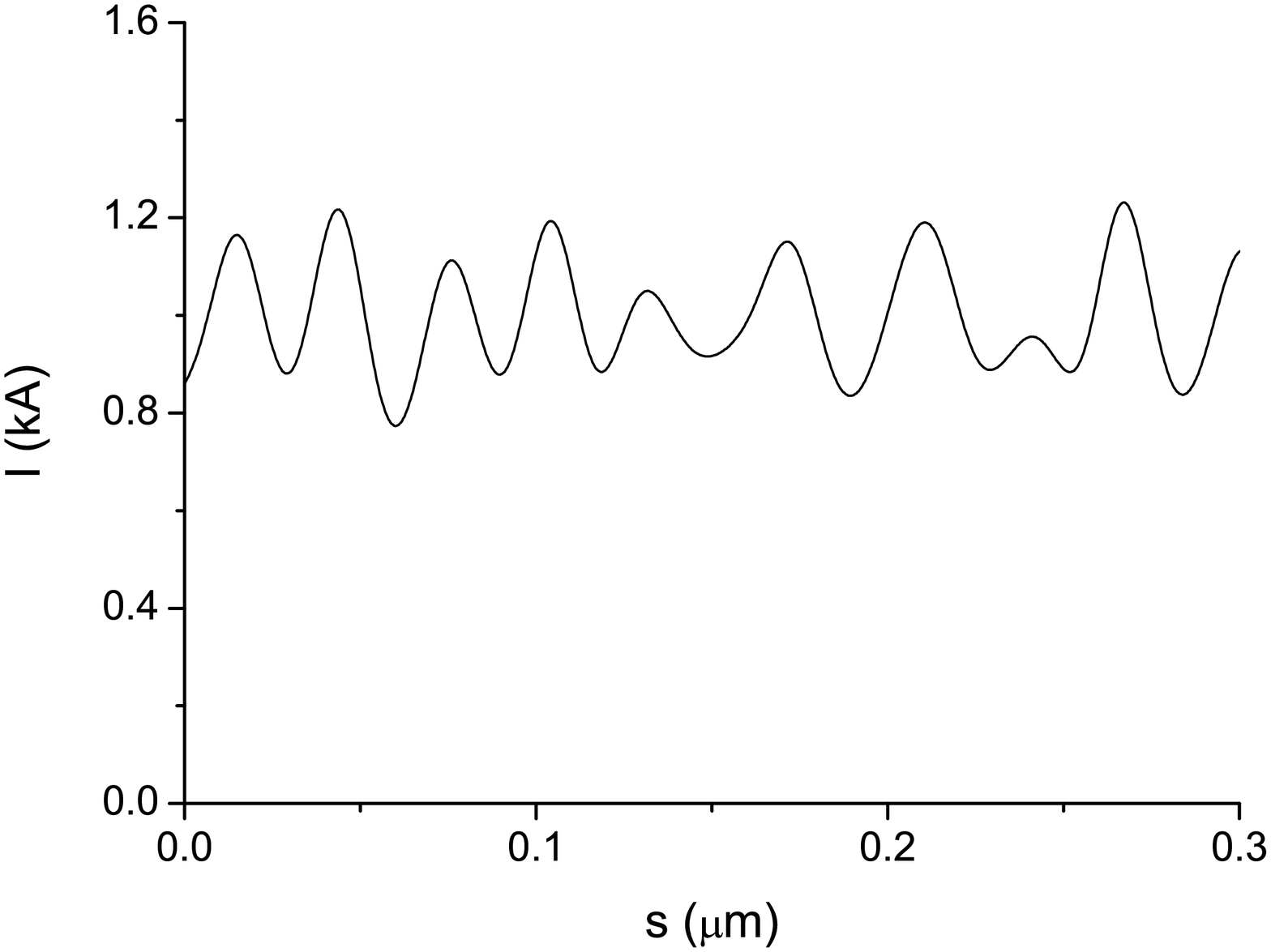}

\caption{\small Modulated electron bunch (left) and its enlarged fraction (right) after the 3rd amplification
cascade. The calculations were performed with LoSCA.}

\label{fig:losca-3rd-time}
\end{figure*}

\begin{figure*}[tb]

\includegraphics[width=0.49\textwidth]{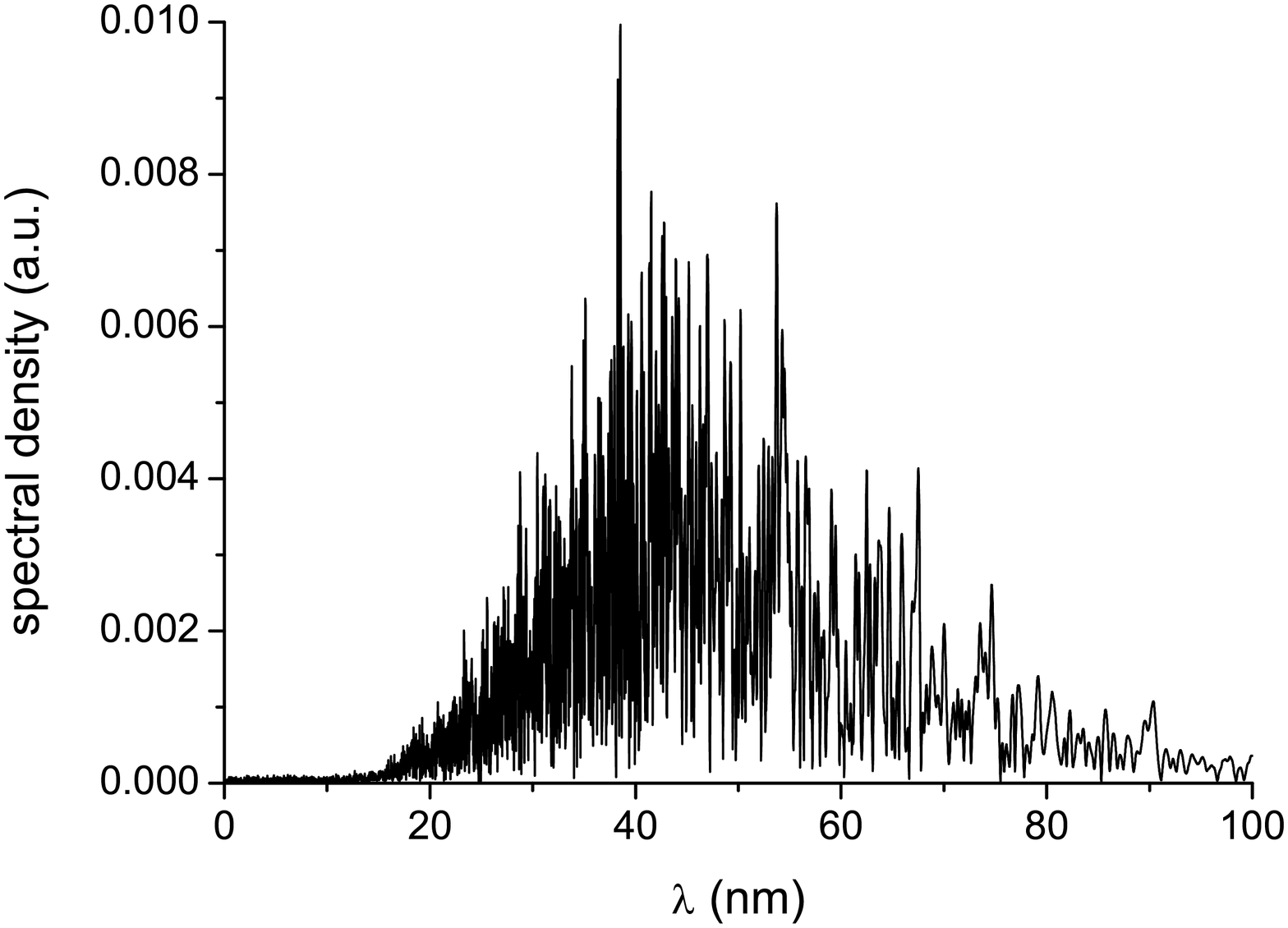}
\includegraphics[width=0.49\textwidth]{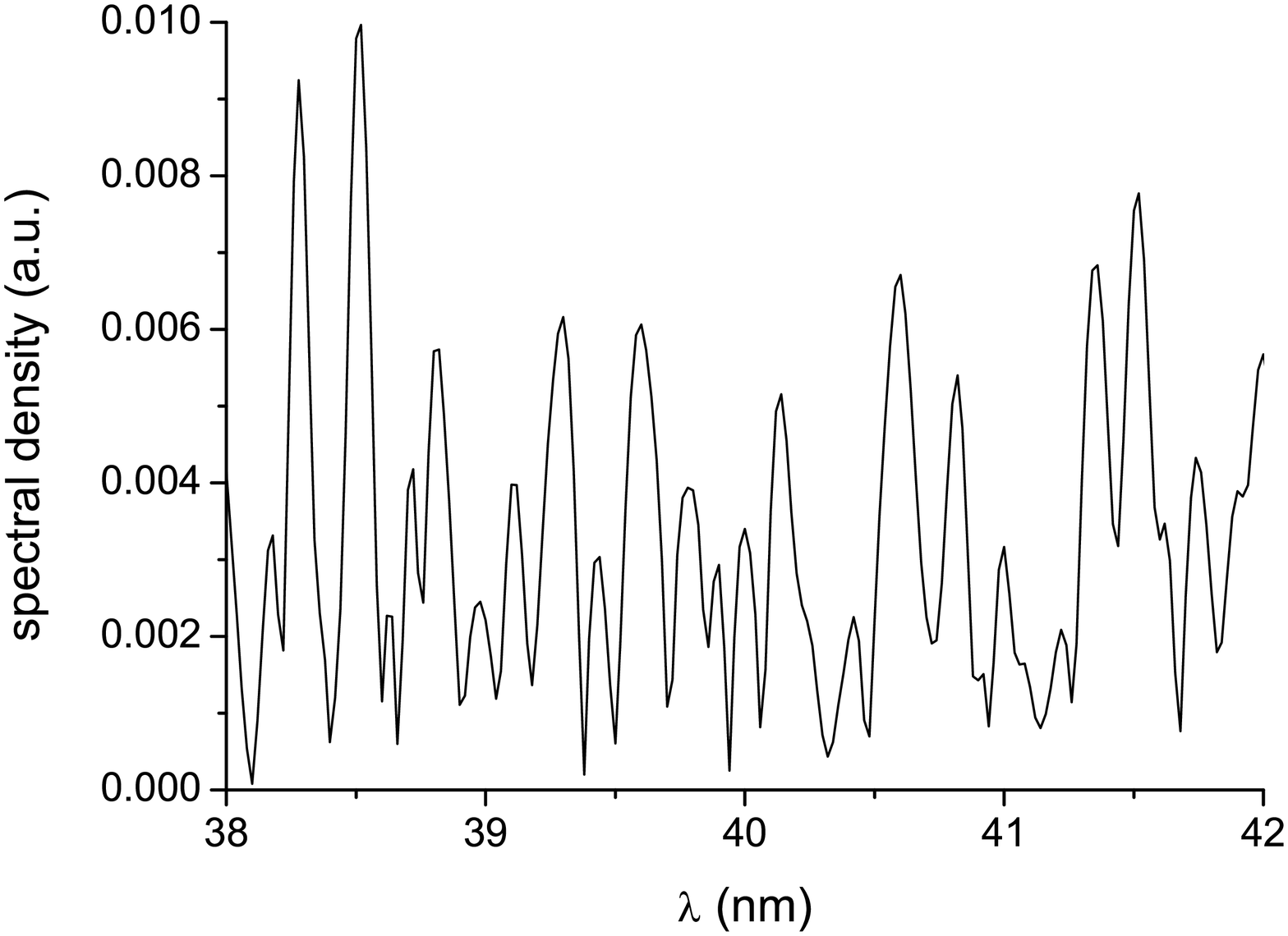}

\caption{\small Spectrum of the modulated electron current (modulus of the bunch form factor) and its
enlarged fraction
after the 3rd amplification cascade. The calculations were performed with LoSCA.}

\label{fig:losca-3rd-spectr}
\end{figure*}

Alternatively, we have developed a simple and fast 1-D code LoSCA ({\bf Lo}ngitudinal {\bf S}pace {\bf C}harge
{\bf A}mplifier). Initially, particles are distributed randomly in the phase
space time-energy. Beam is then cut into short slices (much shorter than
$\lambda_0/2 \pi$), and the LSC interaction between the slices in a drift space is
simulated with the help of the LSC wake function,
averaged over beam cross-section \cite{geloni-wake}.
The drift is divided into pieces, and at the end of each piece the paricles change their positions depending
on their energies. The change of particles' distribution through a chicane is simply performed
by applying $R_{56}$. Note that in our particular case the change of particles' positions happens only in
chicanes. In the drifts it is negligibly small because of a very small spread of the longitudinal velocities
due to both energy spread and emittance, and also because the length of a drift is much smaller that the
reduced wavelength of plasma oscillations \cite{lsca-prst}. This means that for our parameter set
it is sufficient to calculate only energy change of particles due to LSC interaction without dividing the drift
into pieces. In fact, due to a simple model, used in LoSCA, we were able to relatively quickly calculate
the evolution of particles' distribution through LSCA cascades with real number of particles in a bunch (with
the charge of 100 pC),
namely with $6 \times  10^8$ particles. To have a representative set of data, we performed several
simulation runs with different initial shot noise realizations.
The results of simulations of the first three cascades of LSCA
for one of realizations are presented in Figs.~\ref{fig:losca-3rd-time} and \ref{fig:losca-3rd-spectr}.

It is interesting to observe (see right plots in Figs.~\ref{fig:astra-time}  and \ref{fig:losca-3rd-time})
that LoSCA gives the same amplification of shot noise modulations as Astra,
despite the fact that essentially different models are used in these two codes. We propose the following
explanation. For a considered parameter range a typical transverse size of the LSC field is comparable to
the electron bunch transverse size. Particles in Astra execute betatron oscillations and sample different
LSC fields, while in LoSCA there is no transverse motion but the applied LSC wake is averaged over electron beam
cross-section.
Our conclusion is that both approaches lead to very similar results for the LSCA gain,
so that a simplified approach is justified for a typical parameter range of LSCA \cite{lsca-prst}:
amplified wavelength
is in the range described by Eq.~(\ref{opt-wl}), and beta-function is smaller than a length of a drift space in
an amplification cascade.

\begin{figure*}[tb]

\includegraphics[width=0.49\textwidth]{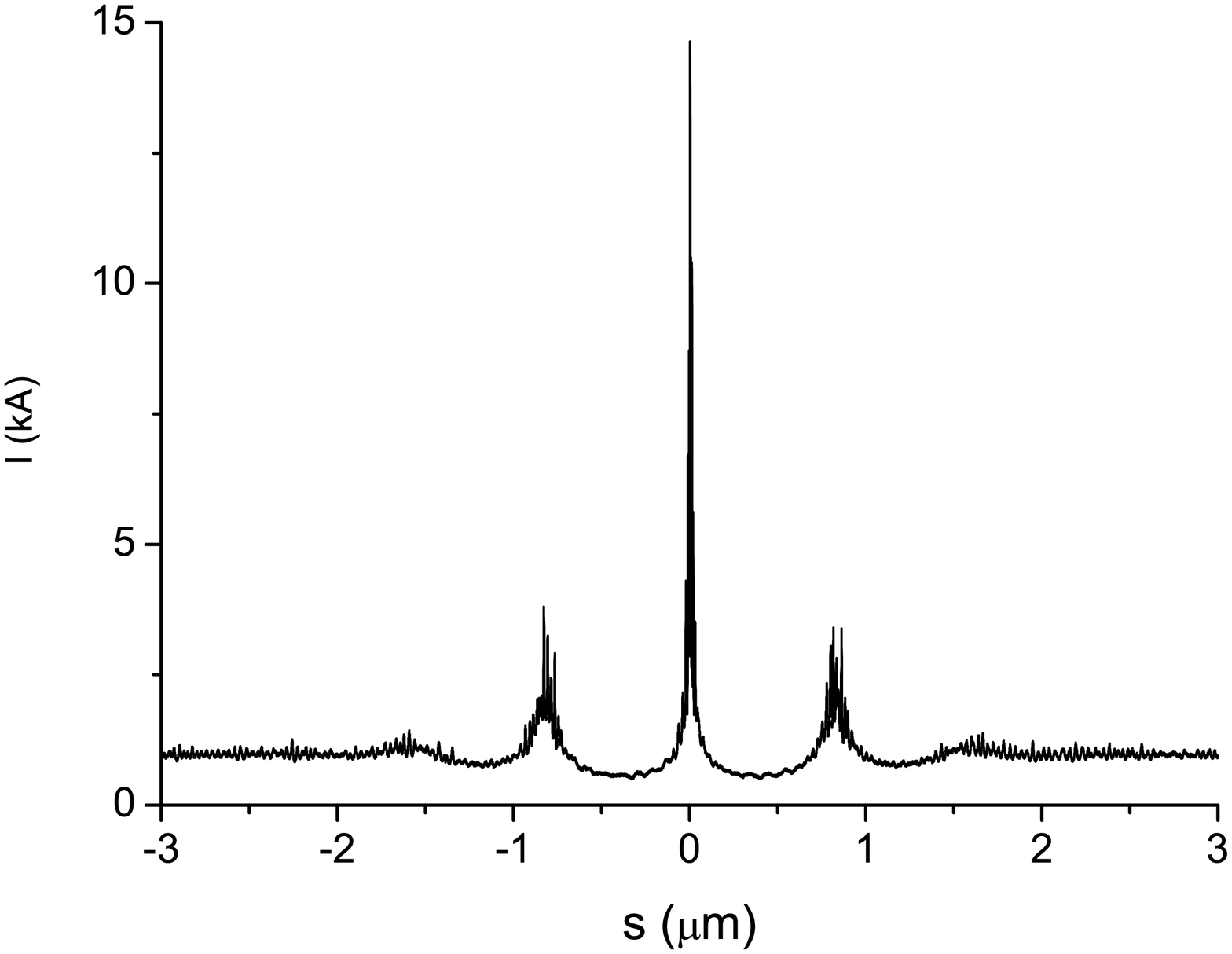}
\includegraphics[width=0.49\textwidth]{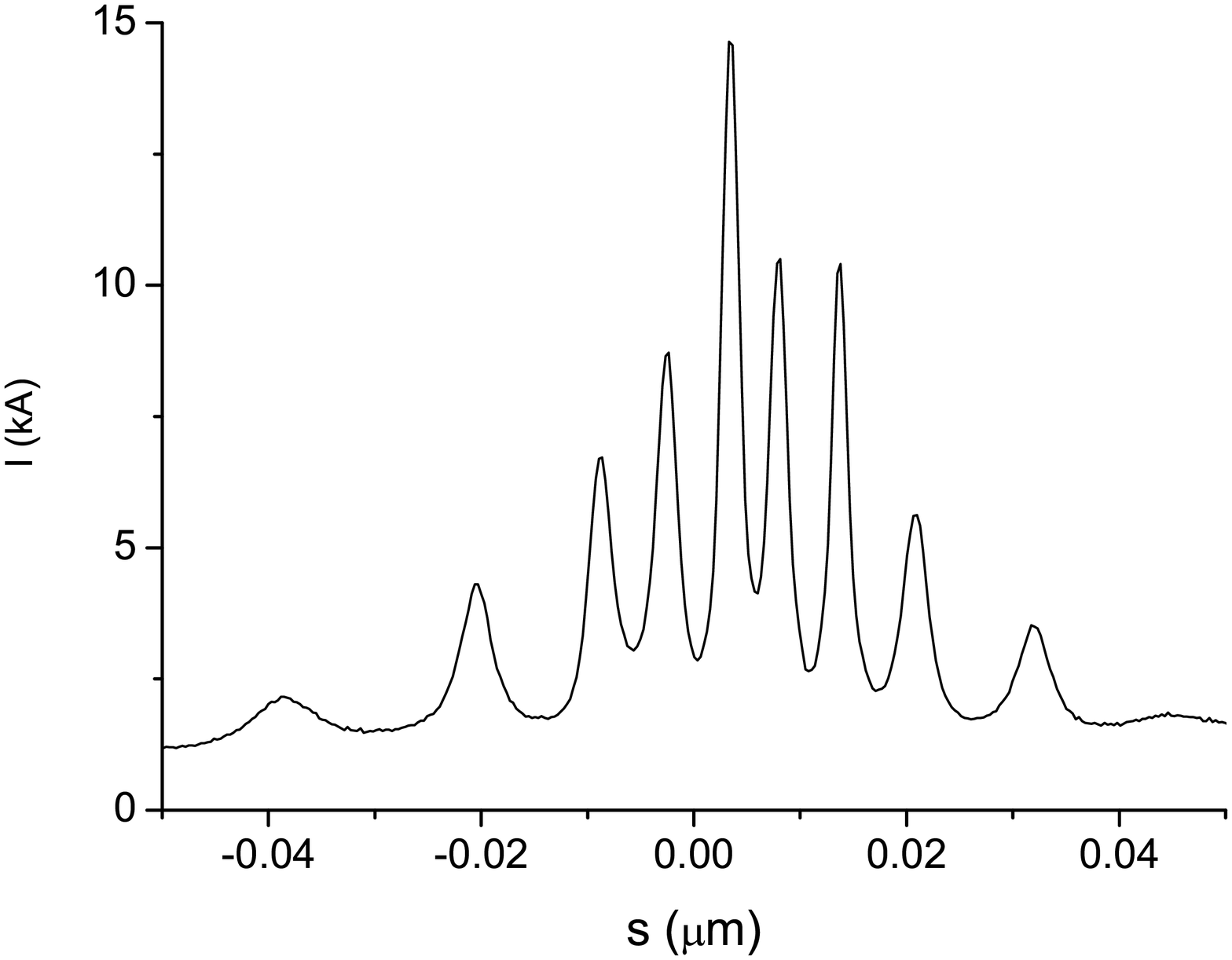}

\caption{\small  Central part of the electron bunch (left) and a zoomed high-current part (right)
after compression in the 4th chicane. The calculations were performed with LoSCA.}

\label{fig:losca-comp}
\end{figure*}

\begin{figure*}[tb]

\includegraphics[width=0.49\textwidth]{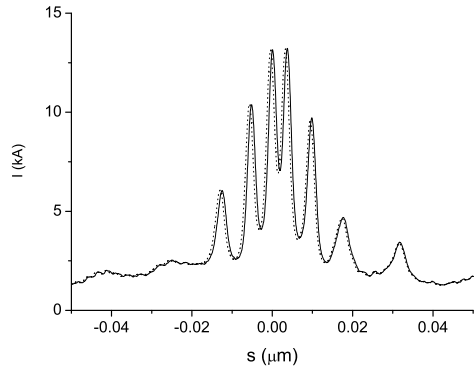}

\caption{\small Results of simulations with CSRtrack.
Zoomed high-current part of the bunch after the 4th chicane for the cases when CSR is off (solid line) and
on (dots). The input distribution was simulated with Astra (but the number of particles was reduced in CSRtrack
simulations).}

\label{fig:csr-nocsr}
\end{figure*}

In the forth amplification cascade the beam first moves in the 2.8 m long drift,
accumulating energy modulation due
to LSC field, and then it is modulated in energy by a short laser pulse.
Modulation of the beam by a laser pulse in two-period undulator was simulated in the same way as it
was done in \cite{attofel-oc,oc-2004-2,prstab-2006-2}, namely by a direct integration of equations of motion
in the combined field of laser and undulator. The resulting energy modulation is presented in
Fig.~\ref{atto-mod}, it was imposed on the phase space distribution of the particles coming from the last drift.
Then the effect of the last chicane was simulated by applying $R_{56}= 7.1 \ \mu$m. A typical current distribution
for the beam, simulated with LoSCA, is presented in Fig.~\ref{fig:losca-comp}. Of course, amplitudes and
positions of peaks (right plot in Fig.~\ref{fig:losca-comp}) are changing shot-to-shot.
A similar distribution is obtained from the simulation with Astra, see Fig.~\ref{fig:csr-nocsr}.

The next problem is a possible influence of
coherent synchrotron radiation (CSR) field in the chicane on
extremely short high-current spikes.
We have performed simulations with the code CSRtrack \cite{csrtrack1,csrtrack2}. We took
the particle distribution at the exit of the forth drift (simulated with Astra), imposed energy modulation
by a laser, reduced number of particles (still having representative ensemble), and
transferred this distribution into CSRtrack. Finally, we
did tracking through the chicane, taking into account CSR wake with the help of the
projected model \cite{csr-projected,martin-csr}. The results can be seen in Fig.~\ref{fig:csr-nocsr}.
As expected \cite{lsca-prst},
no noticeable effects on longitudinal and transverse dynamics were observed (except for a little time shift).
The reason for this is a practically complete
smearing of microstructures inside the chicane due to the coupling of transverse and longitudinal
dynamics described by $R_{51}$ and $R_{52}$ elements of transfer matrix \cite{we-micr,stup-micr,kim-micr}.

\begin{figure*}[tb]

\includegraphics[width=0.49\textwidth]{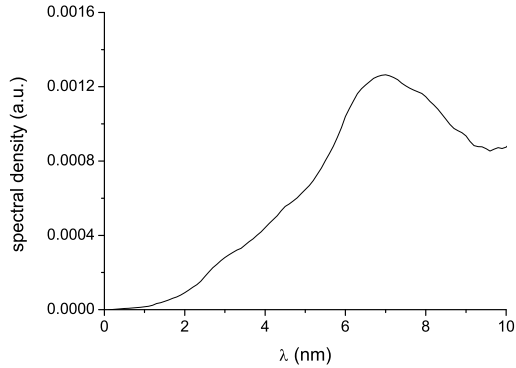}

\caption{\small Ensemble-averaged spectrum of the electron current (modulus of the bunch form factor)
after the last chicane. The calculations were performed with LoSCA.}

\label{fig:losca-comp-spectr}
\end{figure*}

Ensemble-averaged spectrum of the beam current (modulus of the bunch form factor) after the last chicane is presented in
Fig.~\ref{fig:losca-comp-spectr}. Within the spectral window, shown in this Figure, the spectral components
are dominantly coming from the fine structure within the main peak in Fig.~\ref{fig:losca-comp} (right).
The fine structures within two side peaks (which are significantly
weaker compressed), contribute to the spectral density for wavelengths longer than 10 nm.
If a resonance wavelength of the radiator undulator is within the range of 2-8 nm, one would get
quite powerful radiation generated by fine structure within the main peak, i.e. very short pulses - depending
also on the number of undulator periods. Some small contribution from side peaks is, in principle, possible due
to a nonlinear harmonic generation mechanism during LSC amplification and compression.

\begin{figure*}[tb]

\includegraphics[width=0.49\textwidth]{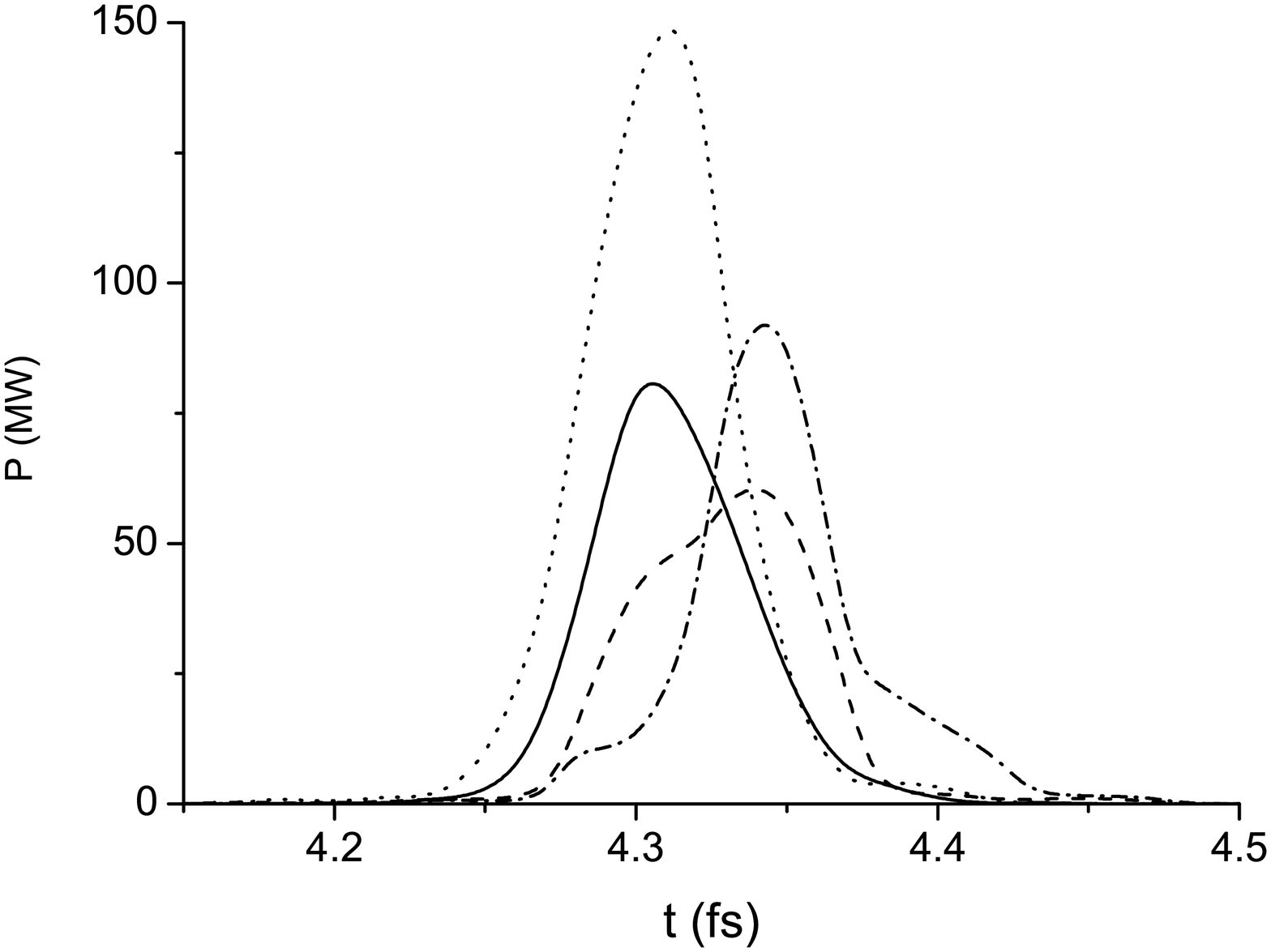}
\includegraphics[width=0.49\textwidth]{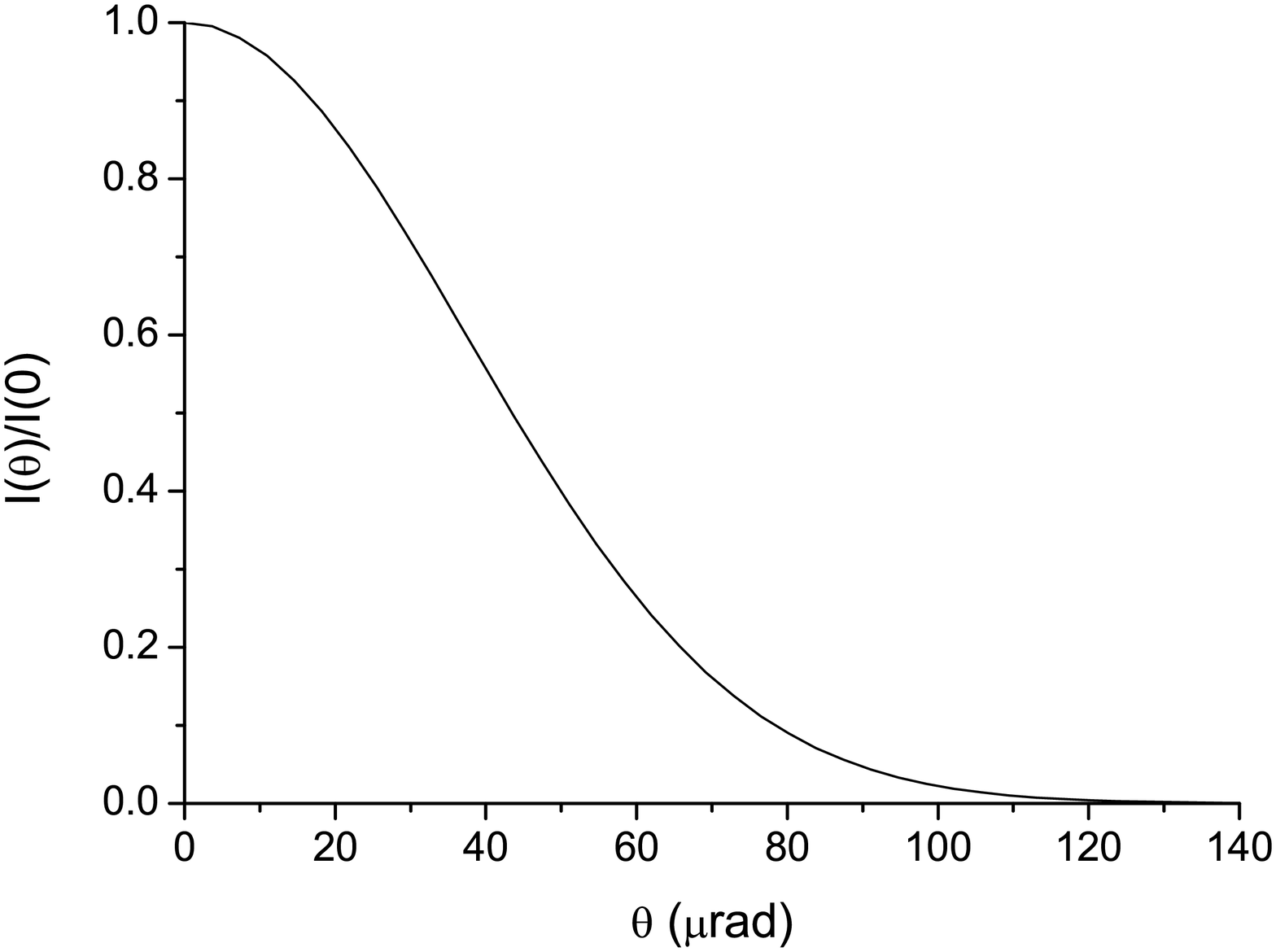}

\caption{\small The results of simulations of the radiation process in the last undulator with the code FAST
when the undulator is tuned to 5 nm.
Left: different realizations of attosecond pulses.
Amplification in LSCA was simulated with Astra (solid) and LoSCA (dash, dot, and dash-dot).
Right: intensity distribution in far zone.}

\label{fig:power-4shots}
\end{figure*}

To illustrate the main properties of the radiation, we assume that the undulator is tuned to
5 nm, and the number of periods is 5 (undulator parameters are presented in the previous Section).
The radiation process was
simulated with the code FAST \cite{fast}. For this purpose, the particles' distribution, simulated
with the help of Astra or LoSCA, was transferred into FAST input distribution.
We should note that FAST, as well as other FEL codes, uses resonance
approximation, i.e. it deals essentially with narrow-band signals. The question arises wether or not one
can properly simulate a process if the input signal (density modulation) has a broad band. We can answer
this question as follows (see \cite{book} for more details): within the central cone of
the undulator radiation the code produces correct results even if the incoming density modulation has a wide
band (in particular, if there is only shot noise having white spectrum). The accuracy of simulations
of radiation properties within the central cone is
on the order of inverse number of undulator periods, i.e. about 20 \% in our case.
Main simulation results are presented in Fig.~\ref{fig:power-4shots}.
Several typical realizations of attosecond
pulses are shown, that were obtained from incoming particles' distributions, simulated with Astra (solid line)
and LoSCA. A typical duration of these pulses is 50-70 attosecons (FWHM), the peak power is at 100 MW level,
and the bandwidth is about 20\%.
Ensemble-averaged pulse energy is 5 nJ with the pulse-to-pulse rms fluctuations about 35 \%.
One can sometimes observe side peaks (corresponding to those in density modulation in
Fig.~\ref{fig:losca-comp}, left plot), but they are typically below 1\% of the power in the main peak.
The rest of the bunch radiates only spontaneously (amplified density modulations in the range
20-80 nm could have produced radiation into large angles, well beyond the central cone,  but this radiation is
totally suppressed due to a finite transverse size of the beam). As one can see from Fig.~\ref{fig:losca-comp},
right plot, the half-angle of coherent radiation is about 50 $\mu$rad. Selecting (with the help of a
pinhole) a cone with half-angle 100-120 $\mu$rad would mean that one does lose not any power
of attosecond x-ray pulses, while spontaneous radiation background is reduced to the level of 50 pJ for
100 pC bunch. Thus, we can state that a contrast (defined as a ratio of radiation energy of
attosecond pulses to the total pulse energy) is high, above 98 \%.

\section{Implementation of the attosecond scheme at FLASH}

\begin{figure*}[tb]

\includegraphics[width=1.\textwidth]{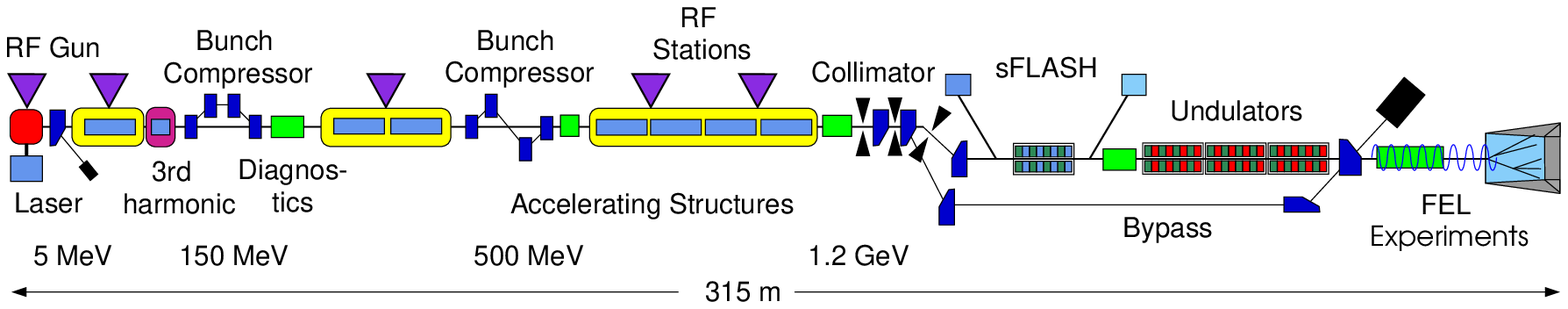}

\caption{\small FLASH layout (http://flash.desy.de).
The laser lab and the experimental hutch are shown as blue
squares.}

\label{flash}
\end{figure*}

FLASH (Free electron LASer in Hamburg) is the soft x-ray FEL user facility \cite{flash-nat-phot,njp}
operating in the wavelength range from 4 to 50 nm. Short VUV and soft x-ray pulses (down to 10 fs
\cite{flash-nat-phot}) are produced in a 27 m long undulator in self-amplified spontaneous emission (SASE)
mode \cite{ks-sase}. Recently, radiation at water window wavelengths down to 4 nm was generated.
FLASH layout is shown in
Fig.~\ref{flash}. The beam is produced by the laser-driven RF gun, compressed in two bunch compressors, and
accelerated in superconducting RF modules up to 1.2 GeV. Then the beam is guided through the dogleg (where it
is collimated),
moves about 50 m towards SASE undulator, radiates there, and goes to the beam dump
(alternatively, it can go to the dump via bypass).

Presently, there are two temporal experimental setups in the straight section between the
dogleg and the undulator.
Optical replica synthesizer (ORS) \cite{we-ors,angelova} is placed (not shown in Fig.~\ref{flash})
directly behind the dogleg/collimator. The ORS is followed by sFLASH \cite{sflash},
the experimental setup for seeding the FEL with the radiation produced by the generation of higher harmonics
of Ti:Sa laser in rare gases. The laser beam together with high harmonics is sent from the laser laboratory
to the tunnel, where it is in-coupled to the electron beam line through the dogleg. In the case of
successful lasing of sFLASH, the amplified VUV radiation is supposed to be outcoupled by a mirror and
transported to the experimental hutch
(at the same time laser beam for pump-probe experiments is transported there
from the laser laboratory). The distance between the dogleg and the mirror is about 30 m.

\bigskip

\begin{figure*}[tb]

\includegraphics[width=0.49\textwidth]{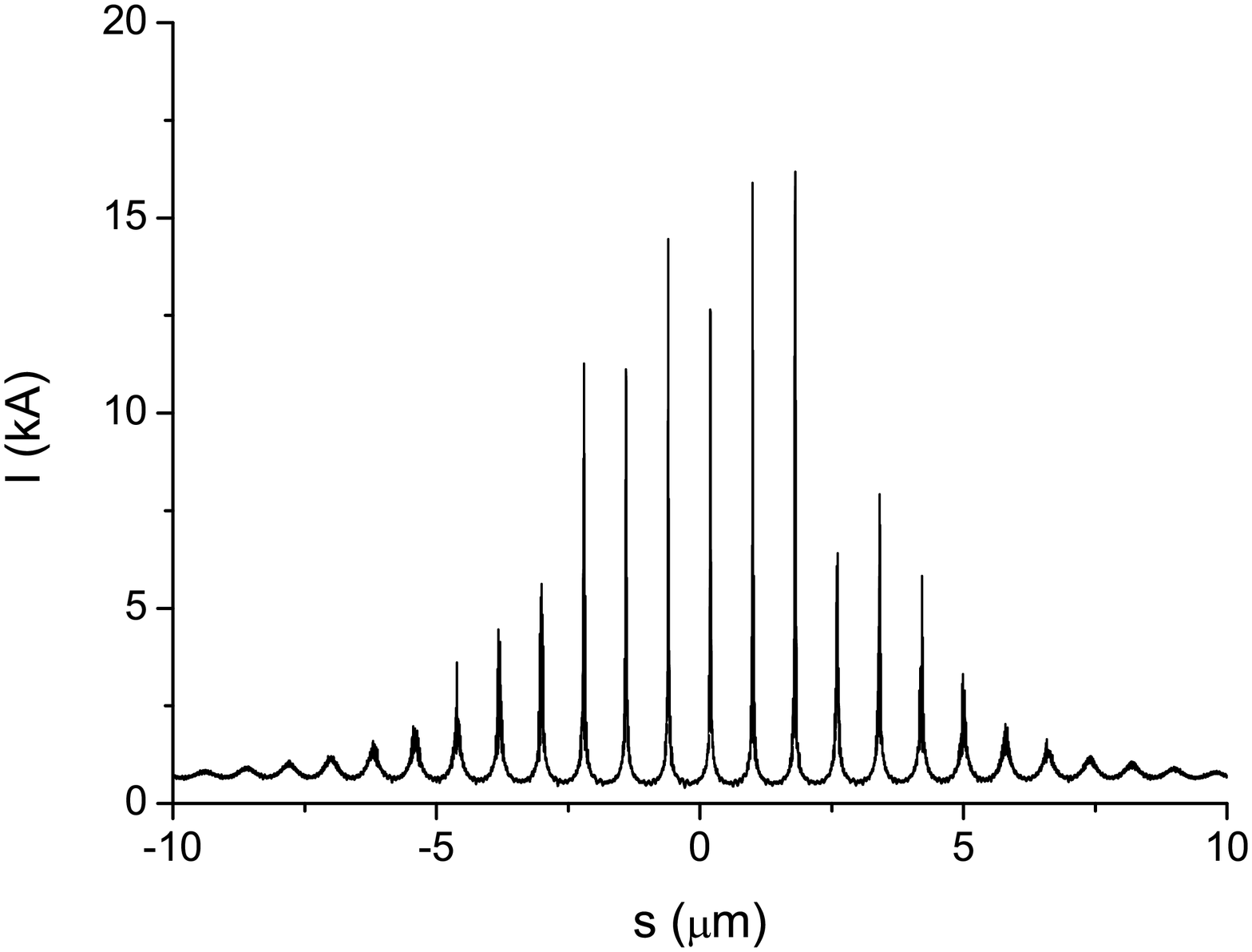}
\includegraphics[width=0.49\textwidth]{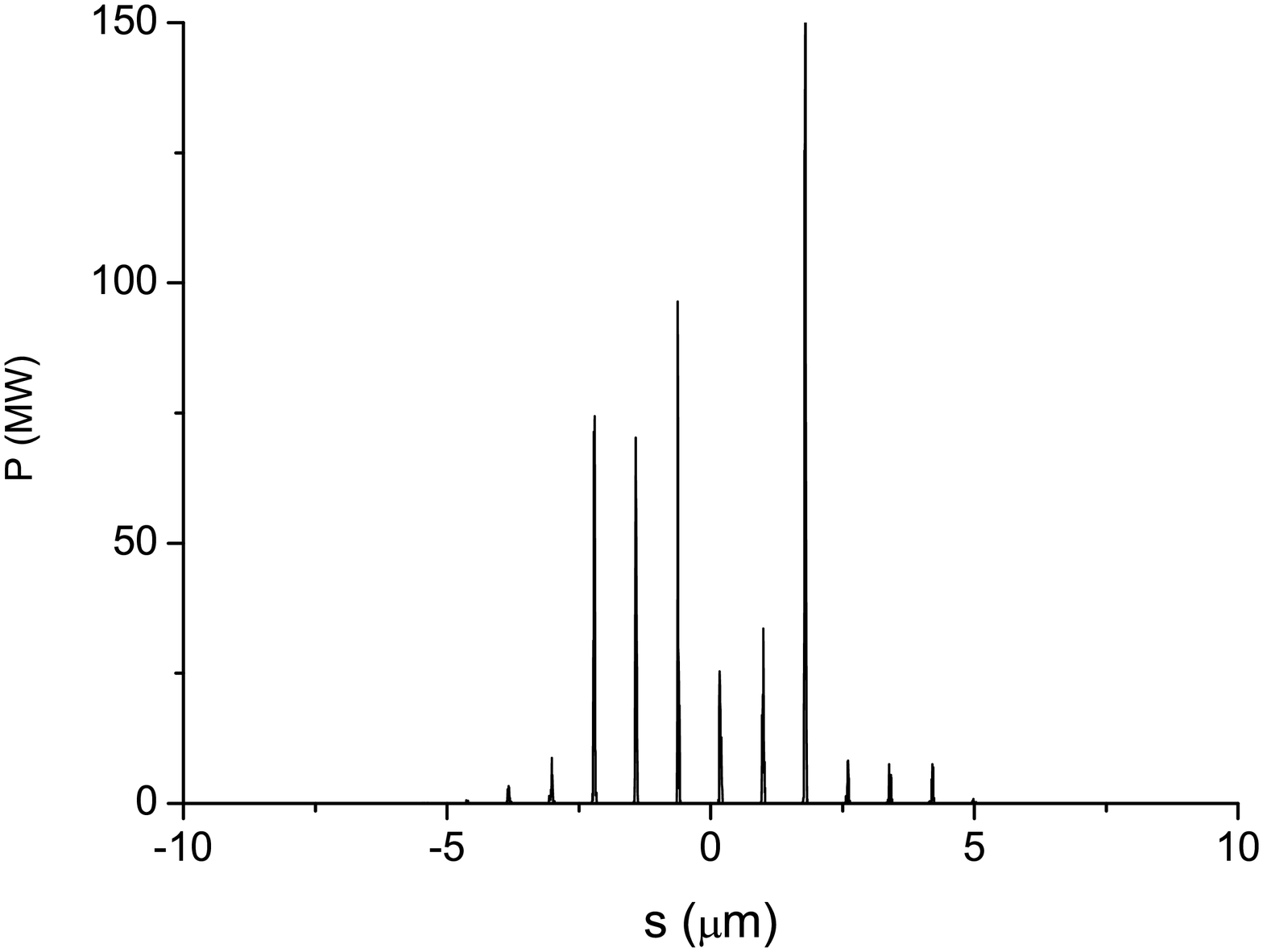}

\caption{\small  Modulated electron bunch at the exit of the 4th chicane (left) and power of the radiation
from the undulator (right) for the case of a relatively long pulse (35 fs FWHM) from Ti:Sa laser.
The calculations were performed with LoSCA and FAST.}

\label{fig:long-pulse}
\end{figure*}

We note that this infrastructure is well suited for production of attosecond pulses in the LSCA and using
them for experiments in the hutch.
The total length of LSCA is supposed to be about 14 m, so that it can
substitute one of the two existing setups. Experiments can be started with
the existing Ti:Sa laser (that delivers
35 fs long pulses with a pulse energy of up to 50 mJ \cite{sflash}) for modulation
of the electron beam, in this case one will produce a train of attosecond soft x-ray pulses.
We have simulated this situation with LoSCA and FAST, the results are presented in Fig.~\ref{fig:long-pulse}.
To get required amplitude of energy modulations, we used only 15 mJ out of available laser pulse energy.

Later the laser can
be upgraded in order to produce 5 fs long pulses with pulse energy about 3 mJ what would allow to obtain
an isolated attosecond x-ray pulse.
The attosecond pulses with the help of the outcoupling mirror
can be transported to the existing experimental hutch where they can be used together
with synchronized few femtosecond long powerful laser pulses.

As a possible option one can consider seeding FLASH undulator with an attosecond pulse. In this case
the outcoupling mirror is moved out, and the chicane at that position is used to slightly delay
the electron bunch so that the radiation pulse is parked on the fresh part of the
bunch. Despite the transverse size of the radiation at the entrance of FLASH undulator is much larger than
that of the electron beam, an effective seed power is still much larger than an effective power of shot noise.
During amplification the radiation pulse would be stretched, so that finally pulse duration would be
comparable to FEL coherence time (about 1 fs).

Finally we note that the attosecond scheme, described in this paper, can also be used at FLASH II \cite{flash2},
FERMI@Elettra \cite{fermi} and similar facilities,
as well as at soft x-ray beamlines of large x-ray FEL facilities like SASE3 beamline
of the European XFEL \cite{euro-xfel-tdr}.

\end{document}